# Imaging the interface of epitaxial graphene with silicon carbide via scanning tunneling microscopy

G. M. Rutter,[1] N. P. Guisinger,[2] J. N. Crain,[2] E. A. A. Jarvis,[2] M. D. Stiles,[2] T. Li,[1] P. N. First,[1,*] and J. A. Stroscio[2,*]

[1] School of Physics, Georgia Institute of Technology, Atlanta, GA 30332
[2] Center for Nanoscale Science and Technology, NIST, Gaithersburg, MD 20899

Graphene grown epitaxially on SiC has been proposed as a material for carbon-based electronics. Understanding the interface between graphene and the SiC substrate will be important for future applications. We report the ability to image the interface structure beneath single-layer graphene using scanning tunneling microscopy. Such imaging is possible because the graphene appears transparent at energies of ± 1 eV above or below the Fermi energy ($E_F$). Our analysis of calculations based on density functional theory shows how this transparency arises from the electronic structure of a graphene layer on a SiC substrate.

To appear in Phys. Rev. B

Graphene, a single layer of $sp^2$-bonded carbon atoms, is an almost ideal two-dimensional system that exhibits several extraordinary transport properties.[1] These include unusual magnetotransport phenomena such as a nonzero Berry phase in the integer quantum Hall effect[2,3] and antilocalization.[4,5] In addition, the high mobility and low dimensionality of graphene make it an attractive material for the development of novel nanoscale electronics. The potential for applicable graphene-based electronics rests on both device performance and the ability to fabricate uniform structures on large length scales reliably and cost effectively. Unlike exfoliation techniques, graphene grown epitaxially on SiC offers a realistic solution for large-scale fabrication and patterning of graphene structures.[6,7]

Optimal performance in graphene-based devices depends on its high mobilities and long carrier lifetimes, which result from the inhibited backscattering due to graphene's symmetries.[8,9] However, the substrates that support graphene structures may break the ideal symmetries or dope the graphene with extrinsic charge, either of which profoundly affects the electrical transport.[4,7,10,11] Interface electronic states that do not contribute to transport directly could affect device operation through electrostatic screening of the external potential used to modulate the graphene carrier density. In particular, one challenge for the graphene-SiC approach is the limited understanding of the interface's influence on the electronic properties and charge transport. To explore the role that the substrate plays in this graphene system, we combine atomic resolution measurements via scanning tunneling microscopy (STM) and spectroscopy (STS) along side electronic structure calculations to characterize the interface between graphene and its SiC substrate.

In this paper, we first describe our preparation of graphene layers on SiC and give some details of the scanning tunneling microscope we use to measure these systems. After that background, we discuss the three different thicknesses of graphene observed in our samples and the structural and electronic properties that distinguish them. Then, we discuss the interface structure that is observed for thin graphene layers at high bias voltages. These measurements allow us to interpret the reconstructions observed in previous measurements. Following these experimental observations, we describe the electronic structure calculations we performed to provide additional insight into the phenomena giving rise to our STM images. The calculations show that the large density of states at the interface and the hybridization between those states and the graphene give rise to the apparent observation of the interface structure above the surface for voltages outside the SiC band gap. Finally, the calculations suggest an alternate interpretation for parts of the surface that appear free of graphene as a strongly interacting carbon layer.

We prepared epitaxial graphene on semi-insulating 4H-SiC(0001) samples by thermal desorption of silicon at high temperatures.[6,12] The sample was first hydrogen etched and then annealed to temperatures above 1200 °C, where graphitization occurs at the SiC surface. The graphene thickness can be controlled by the temperature, and to a lesser degree by the annealing time, allowing the preparation of samples with varying thicknesses in the range of one to three layers. The majority of data reported in the present study was acquired on samples with an estimated average thickness of one layer, as determined by low-energy-electron diffraction (LEED) and Auger electron spectroscopy. STM experiments on the graphene films were performed in a custom-built cryogenic ultrahigh vacuum instrument. All measurements reported here were acquired at a temperature of 4.3 K using Ir probe tips. STS measurements were conducted by applying a small 500 Hz modulation to the sample voltage and by measuring the differential conductance, *dI/dV*, with lock-in detection. Results similar to those presented here have been found with different probe tips and samples at temperatures of 4.3 K and 300 K.

Survey images of the first graphene layer on the SiC substrate show the graphene lattice structure superimposed with interesting adatom features (Fig. 1). We assign these adatom features as subsurface interface structures imaged beneath the first graphene layer, a similar conclusion drawn by other groups.[13,14] Single-layer graphene can be identified from a detailed study of the different terraces found on the surface. Three types of terraces are found typically, separated by successive steps up in surface height, as shown in Figs. 2(a) and 2(b). Figure 2(a) shows two terraces of similar appearance, separated by a 2.5 Å step (the upper terrace is the dominant type found on these samples, see Fig. 1). Each terrace has a high density of adatom features, but there are significant differences between the two. A major difference between the upper and lower terraces [Fig. 2(a)] is that the graphene lattice can be imaged at low tunneling biases on the upper terrace as a honeycomb structure (Fig. 1), whereas no graphene lattice is observed on the lower terrace. A natural interpretation of these results is that the lower terrace is the reconstructed SiC substrate without graphene (layer 0), and the upper terrace is the first graphene growth layer (layer 1).

Support for this interpretation arises from differential conductance measurements of layer 0 and layer 1, which reveal a metallic or semimetallic spectrum on the upper terrace, whereas the lower terrace has a 300 mV band gap around $E_F$ [Fig. 2(c)].

However, for single-layer graphene, the density of states should go to zero at the point where the electron and hole bands meet, called the Dirac point. This reduction in the density of states should be observable in scanning tunneling spectra. While previous measurements of the band dispersion have determined the Dirac point to be around 300 meV below $E_F$,[15,16] a reduction in the density of states at the Dirac point is not observed in the spectra of single-layer graphene. A plausible explanation for the absence of this signature is the added contribution from the interface to the density of states at energies away from $E_F$, as will be elaborated later.

Measurements at low tunneling bias show large differences between layers 1 and 2. In particular, there are dramatic differences in the topography [Fig. 2(b)] comparing layer 2 to layer 1. Instead of a terrace dominated by adatom type features, the graphene lattice is the dominant topographic feature, and is easily observed with atomic scale resolution.[16] Used as a fingerprint, all these STM characteristics allow for a unique local identification of single-layer (layer 1) and bilayer (layer 2) epitaxial graphene.

STM images of the first graphene layer obtained at different tunneling biases (Fig. 3) show that this layer appears transparent to tunneling at energies well above or well below $E_F$, as has been observed by others.[13] Adatom features dominate the images for both unoccupied and occupied states at biases of ± 1 V. From empty-state images [e.g., Fig. 3(a)], we identify two predominant adatom structures: pyramidal clusters and hexagonal rings. These structures closely resemble Si adatom structures observed in the reconstruction of bare SiC and Si surfaces.[17-20] The first features, pyramids, resemble structures observed by annealing a Si-rich surface of SiC(0001) to form a 3x 3 reconstruction, which has been described by the Starke model.[19,20] Within this model, four Si adatoms arrange in pyramidal clusters (tetramers), which are similar to features observed in Figs. 1, 3(a), and 3(b). At high voltage, the tetramers appear as one object, but at lower tunneling bias the three bottom adatoms of the tetramer become visible [Figs. 1 and 3(b)]. When imaging the filled states, the top adatom of the tetramer appears transparent, resulting in a trimer structure [Fig. 3(e)]. From the bias dependence, it is apparent that the tetramers play a key role in the graphene morphology. Figures 3(a)-3(c) show that there is a direct correspondence between the Si tetramer features and "6x6" maxima in the graphene dominated images. This indicates that the 6x6 periodicity observed in graphene layers grown on SiC is due to a SiC interfacial reconstruction, and not a moiré effect as previously suggested.[21,22]

The other adatom features, hexagonal rings, resemble corner holes observed within the dimer-adatom-stacking-fault model of Si(111)-7x7.[17] A simplified structural model of the tetramers and the hexagonal rings is shown in Fig. 4(a). These structures are suggested by STM images of the reconstructed SiC surface during graphitization, as seen in Fig 4(b). In the STM images (Fig. 4(b)), the corners of each hexagon fall on a SiC $\sqrt{3}$x$\sqrt{3}$ R30º sublattice, but adjacent hexagons lie on different sublattices. Overlaid on the image of Fig. 4(b) are the three SiC $\sqrt{3}$x$\sqrt{3}$ R30º sublattices (red, blue, and green crosses), which together occupy all of the SiC 1x1 lattice sites. Color-coded circles show the registry of adjacent hexagonal rings with the underlying sublattices. The interface is not perfectly ordered, but areas such as these where adjacent hexagons fall on different sublattices are typically seen in these samples. This interface structure, comprised of equivalent structures on each of the three SiC $\sqrt{3}$x$\sqrt{3}$ R30º sublattices, explains many features of the SiC $6\sqrt{3}$x$6\sqrt{3}$ R30º pattern observed in LEED measurements.[23]

Electronic structure calculations for a single graphene layer on the Si terminated SiC(0001) surface can explain the observation of the graphene lattice at low biases and the interfacial adatom structures at higher biases. We performed all-electron density-functional calculations using a generalized gradient approximation[24] for the exchange-correlation potential and a local numerical basis of double-numeric quality (plus polarization functions on the Si and C in the case of the smaller interface cells).[25] There is not a good small-surface-area lattice match between a graphene sheet and the SiC(0001) surface. Initial interface structures were created using the experimental SiC lattice vectors and uniformly straining the graphite monolayer to accommodate these. The strain inherent in this enforced commensuration of lattices necessary for satisfying the periodic boundary conditions in our calculations was 8% for the √3 SiC interface cell and 0.5% for the 2√3 SiC interface cell. For the 5x5 cell, employed to accommodate the observed adatom structures for the analysis of the iso-wave-function and charge densities, the strain was 0.2%. All atomic coordinates were permitted to relax, subject to periodic boundary constraints, toward a local minimum from their initial starting geometries. Starting geometries with different translations of the graphene layer and different heights of this layer over the SiC substrate were performed for several test cases. Based on the analogy with the bare SiC reconstructions, we assumed that the adatoms were Si. We expect the results to hold if these were replaced with C. Monkhorst Pack grids of 5x5x1 and 9x9x1 were employed for the k-point sampling during the optimization of the atomic coordinates of the √3 SiC interface cells (with similar behaviors observed in both cases), and initial structures were relaxed employing thermal smearing of 0.15 eV. The final densities employed for analysis were taken from a self-consistent field calculation with Fermi occupancy. The iso-wave-function plots are from calculations with only the gamma point included.

A first-principles density-functional theory calculation for a graphene layer above a tetramer with neighboring $T_4$ adatom [boxed region in Fig. 4(a)] gives insight into the transparent nature of imaging the first graphene layer. Figure 5 shows a series of iso-wave-function contours for three different energies (a) below, (b) near, and (c) above $E_F$. SiC interface orbitals dominate the contours for energies above and below $E_F$, in agreement with the experimental findings [Figs. 3, (a) and (d)]. In contrast, graphene states dominate the contours for energies within 0.1 eV of $E_F$, which accounts for the trend toward imaging the graphene lattice at low bias [Figs. 3, (c) and (f)]. A large isocontour value was chosen to highlight the difference between the graphene states at $E_F$ and the apparent gap in the SiC substrate density of states. A smaller isocontour value shows finite graphene density away from $E_F$. Interestingly, this orbital analysis also displays the difference observed in the appearance of the tetramers for filled versus empty states. Specifically, the on-top site of the tetramer has no orbital contribution over the displayed energy range for the filled states [Fig. 5(a)], but is apparent in the empty states [Fig. 5(c)] leading to the appearance of trimers rather than tetramers in the STM images [Figs. 3, (a) and (e)].

These calculations also give insight into the "6 x 6" corrugation observed in the graphene images. Figure 5(d) shows a top view of the atomic positions for the calculated interface, while Fig. 5(e) shows the corresponding total charge density for a slice parallel to the interface positioned just above the graphene layer. We observe qualitatively good agreement between the charge density image and the graphene images [Figs. 3, (c) and (f)]. The larger charge density in the vicinity of the Si tetramers arises from the buckling of the graphene lattice over the Si adatoms. This suggests that the 6x6 corrugation observed in the STM images is largely due to the graphene lattice draping over features of the interface reconstruction. In fact, the experimental corrugation amplitude of ≈ 0.6 Å is the same as the geometric displacement we calculate. This is not surprising since it is a common feature of the graphene lattice to deform and cover surface features.[16]

Another intriguing result from the calculations suggests an alternate interpretation of layer 0 as a nonmetallic carbon layer strongly coupled to the SiC substrate. For an interface structure without adatoms, the calculations (for both the √3 SiC interface cell and the 2√3 SiC interface cell) give two energetically stable configurations for the first carbon layer, depending on the initial graphene-SiC distance prior to full relaxation of the atomic coordinates. The structure discussed above is a weakly interacting graphene layer above the SiC surface, with the graphene sheet 3.4 Å above the SiC surface. The highly interacting structure involves direct bonding interactions between the surface Si and the C in the graphene, and disrupts the planarity of the graphene sheet. The bonds formed between the surface Si and the graphene C are 2.0 Å compared to 1.9 Å within SiC. The surface electronic structure is markedly distinct between these two interface geometries. The highly interacting graphene structure yields a semiconducting gap in the vicinity of $E_F$, similar to that observed for layer 0, whereas the weakly interacting graphene layer shows a metallic density of states, similar to that observed for layer 1. These findings suggest an alternative model for the layer 0 terrace in Fig. 2(a) as consisting of a highly interacting graphene layer as opposed to a bare SiC surface with quasiordered Si adatom structures. If layer 0 were the strongly interacting graphene layer on the substrate and layer 1 were the weakly interacting layer, our calculations would suggest a step height between the two of 1.4 Å. This value is smaller than the observed step height between layers 0 and 1 of 2.6 ± 0.4 Å, but the calculated step height between the weakly interacting layer and the bare substrate, 3.4 Å, is higher by a similar amount. These step heights were calculated in the absence of adatom structures, which could change the results. In addition, calculated structural steps heights are frequently different from step heights observed in STM due to electronic effects in the tunneling process. This analysis of the step heights leaves both possible interpretations of layer 0 plausible. Additional support for the interpretation of layer 0 as a strongly interacting graphene layer comes from photoemission spectra of the initial stages of graphitization on SiC,[26] in which graphene σ bands are observed, but the π bands are absent.

In summary, we have shown that the single-layer graphene on SiC can be identified by bias-dependent STM imaging, which displays a superposition of SiC interface features and the graphene lattice. Calculations based on density-functional theory show that the tunneling transparency of the graphene layer arises from the energy dependence of the density of states. The tunneling transparency of the first layer of graphene allows structural features of the SiC interface to be examined on an atomic scale. These SiC interface structures may play an important role in the transport properties of graphene and remain to be examined with atomic-scale measurements of graphene's transport properties.

We thank Ed Conrad, Walt de Heer and Claire Berger for valuable comments and discussions, and Steven Blankenship, Frank Hess, Alan Band, and Nate Brown for their technical assistance. This work was supported in part by the Office of Naval Research, by Intel Research, and by NSF grant No. ECS-0404084.

*To whom correspondence should be addressed: joseph.stroscio@nist.gov; first@physics.gatech.edu

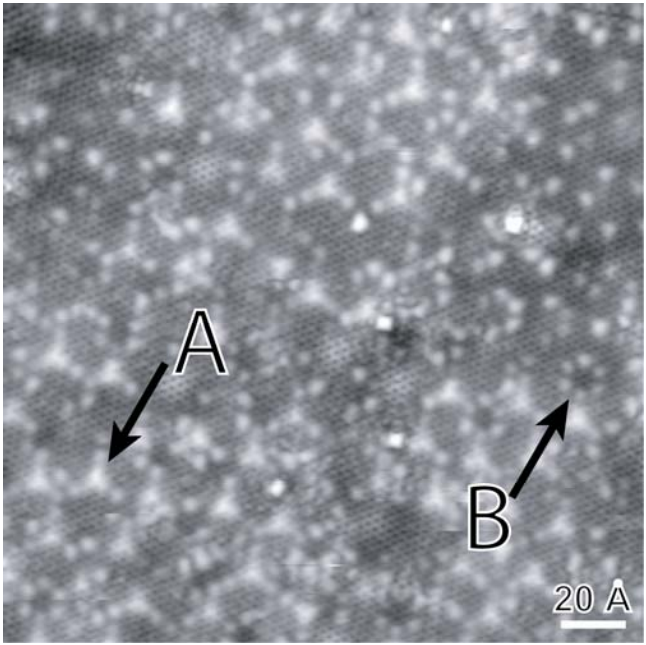

FIG. 1. STM topographic image of the first graphene layer showing a combination of SiC interface features along with the graphene lattice due to the transparency of the graphene ($V_t$=400 mV and $I_t$ = 50 pA). Typical adatom features are tetramers (labeled A) and hexagons (labeled B).

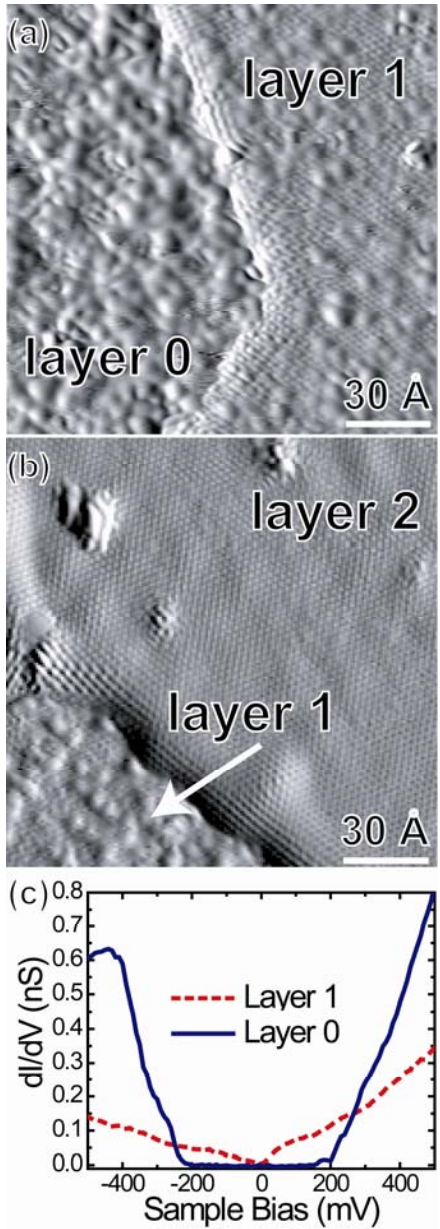

FIG. 2. (color online) (a) STM image showing two regions (layer 0 and layer 1) separated by a 2.5 Å step ($V_t$=600 mV and $I_t$ = 100 pA). (b) STM image showing a 3 Å step up from the first graphene layer to the second layer ($V_t$=300 mV and $I_t$ = 100 pA). The image gray scale is proportional to the horizontal gradient of the topographic height for visual clarity of the two terraces for both images. (c) Differential conductance measurements obtained on the layer 0 and layer 1 terraces.

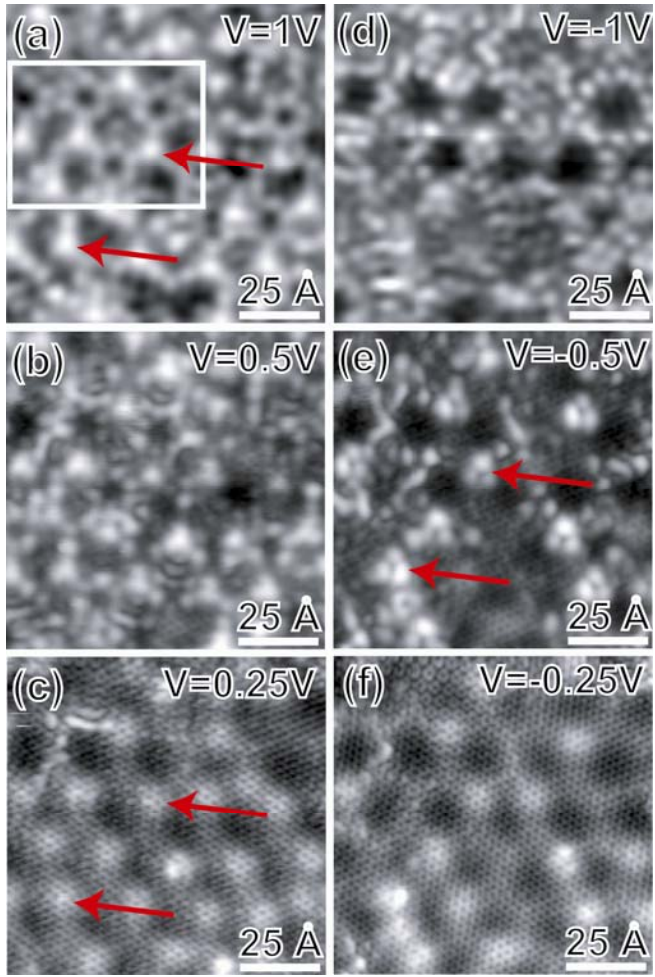

FIG. 3. (color online) Bias-dependent topographic images show the progression from imaging the SiC interface structure at high bias to imaging the graphene overlayer at low bias. The tunneling current is fixed at 100 pA, and the bias voltages are (a) 1.0 V, (b) 0.5 V, (c) 0.25 V, (d) -1.0 V, (e) -0.5 V, and (e) -0.25 V. Red arrows (color online) indicate that different features [tetramers in (a), graphene 6 x 6 maximum in (c), and trimers in (e)] are imaged at the same surface location, dependent on bias voltage. The white box in (a) designates the area magnified in Fig. 4(b).

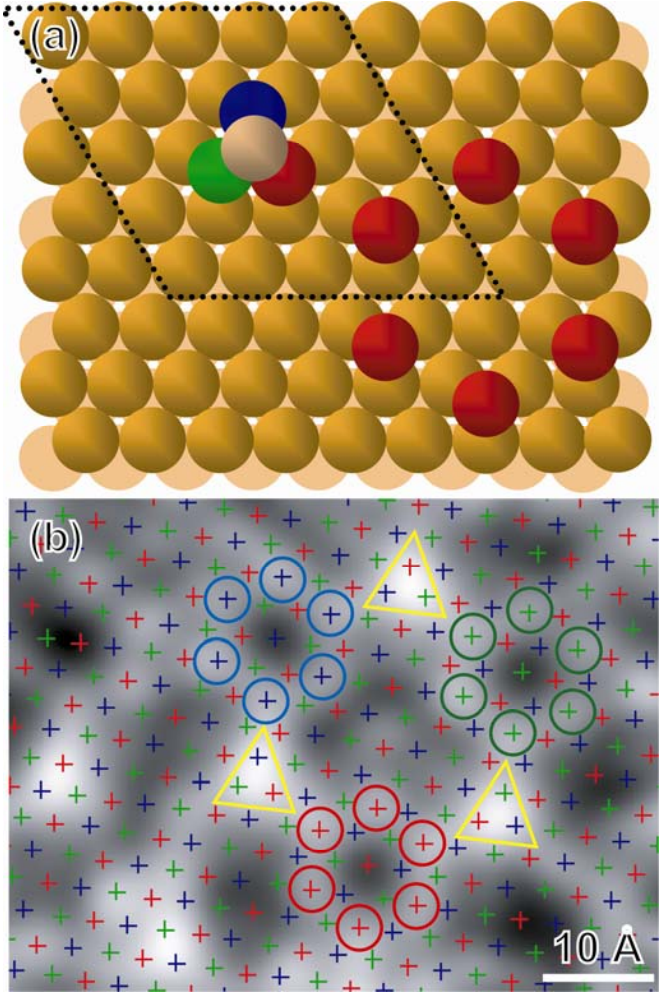

FIG 4. (color online) (a) Schematic geometry of possible Si adatom features consisting of one tetramer and hexagon. The three different colors (red, blue, and green) correspond to Si adatoms on three different sublattices as in (b). The gold atoms represent the Si atoms in the SiC substrate. The region outlined by the dotted cell was used for the calculations described in Fig. 5. (b) Magnified view of the first layer of graphene from Fig. 3(a). Three hexagons are observed to lie on the three different SiC $\sqrt{3}\times\sqrt{3}$ sublattices, denoted by the three different colors. Tetramer features (yellow triangles) are what allow hexagons to switch to different $\sqrt{3}\times\sqrt{3}$ sublattices.

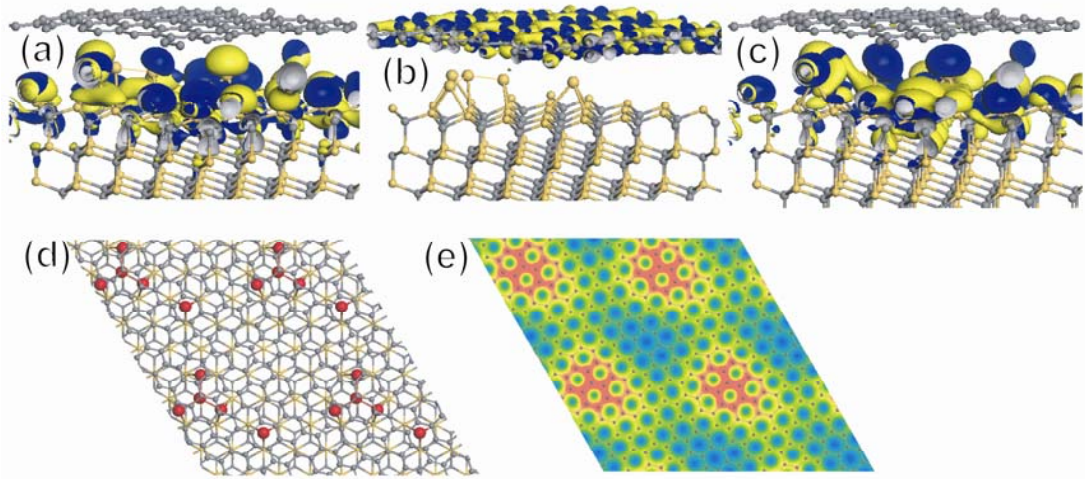

FIG 5. (color online) Iso-wave-function contours for a 5x5 SiC periodic cell with a tetramer and neighboring $T_4$ adatom [boxed region of Fig. 4(a)] with a graphene overlayer. The states are summed over energy windows of (a) roughly -0.8 to -0.1 eV below $E_F$, (b) within ≈ 0.1 eV of $E_F$, and (c) about 0.1-0.8 eV above $E_F$. The color scheme denotes the phase of the orbital. (d) Top-down view of the 5x5 cell (repeated for ease of viewing) with a tetramer and neighboring $T_4$ adatom at the interface displayed in red. (e) Slice of the total charge density above the graphene layer with C atom sites indicated. Here, red indicates regions of highest charge density, and blue corresponds to lowest charge density.